\documentclass[12pt]{iopart}

\usepackage[dvips]{graphicx}
\usepackage[dvips]{color}

\usepackage{iopams}
\usepackage{setstack}
\usepackage{latexsym}
\usepackage{units}
\usepackage{ulem}

\newcommand{\txt}[1]{\mathrm{#1}}

\newcommand{\EC}{E_\txt{c}}

\newcommand{\Ec}{\EC}
\newcommand{\Cj}{C_\txt{j}}
\newcommand{\kB}{k_\txt{B}}

\newcommand{\D}{\,\mathrm{d}}

\newcommand{\GR}{\Gamma_\txt{R}}
\newcommand{\GL}{\Gamma_\txt{L}}

\newcommand{\nS}{n_\mathrm{S}}
\newcommand{\fS}{f_\mathrm{S}}
\newcommand{\fN}{f_\mathrm{N}}

\newcommand{\RT}{R_\mathrm{T}}

\newcommand{\Tenv}{T_\txt{env}}

\newcommand{\IN}{I_\txt{N}}

\renewcommand{\Re}{\mathrm{Re}}

\newcommand{\be}{\begin{equation}} \newcommand{\ee}{\end{equation}}
\newcommand{\ba}{\begin{eqnarray}} \newcommand{\ea}{\end{eqnarray}}

\newcommand{\ie}{i.\,e.}
\newcommand{\eg}{e.\,g.}

\begin{document}

\title[Single-charge escape processes through a hybrid turnstile...]
{Single-charge escape processes through a hybrid turnstile in a dissipative environment}
\author{Sergey V Lotkhov$^1$, Olli-Pentti Saira$^2$, Jukka P Pekola$^2$ and Alexander B Zorin$^1$}
\address{$^1$ Physikalisch-Technische Bundesanstalt, Bundesallee
100, 38116 Braunschweig, Germany}
\address{$^2$ Low Temperature Laboratory, Aalto University, P.O. Box 13500, FI-00076 AALTO, Finland}
\ead{Sergey.Lotkhov@ptb.de}

\date{\today}

\begin{abstract}
We have investigated the static, charge-trapping properties of a hybrid
superconductor---normal metal electron turnstile embedded into
a high-ohmic environment. The device includes a local Cr resistor on one side
of the turnstile, and a superconducting trapping island on the other side. The
electron hold times, $\tau \sim \unit[2-20]{s}$, in our
two-junction circuit are comparable with those of
typical multi-junction, $N \ge 4$, normal-metal single-electron
tunneling devices. A semi-phenomenological model of the environmental
activation of tunneling is applied for the analysis of the switching
statistics. The experimental results are promising for electrical metrology.

\end{abstract}

\submitto{\NJP}

\maketitle

\section{Introduction}

Currently, a growing interest is observed to the quantum sources of
electrical current, based on a clocked transfer of the elementary
charges. Developed for mesoscopic systems of small tunnel junctions,
an orthodox theory of single-electron tunneling (SET)
\cite{AverinLikharev91} has been used to describe the operation of the SET
pumping devices (for a few examples, see
Refs.~\cite{GeerligsTurnstile92,PothierPump,Pekola-NaturePhysics08})
in terms of cyclic sequences of the charge states with the charge
values on the nodes being a multiple of the electron charge $e \approx
\unit[1.602 \times 10^{\txt{-19}}]{C}$. The quantized current value is given by $I = k \times ef$,
where $f$ is the cycling frequency and $k$ is the number of electrons transferred
in one cycle.

A more complete picture involves  a broad scope of cotunneling
leakage processes \cite{AverOdNaz,AverinPekola08} as well as the
influence of the electromagnetic environment \cite{KautzPAT00,PekolaEA10}
to explain the finite, often unexpectedly short decay times of the
participating  charge configurations, imposing a significant limitation
onto the pumping accuracy. Accordingly, the rate of charge leakage
across the whole device has been an important characteristic of any Coulomb
blockade circuit (see, e.g.,
Refs.~\cite{PothierPump,Pekola-NaturePhysics08,KellerScience99}).
For example, in the quantum standard of capacitance, this insulation
property was ultimately quantified by measuring a single-electron hold time of
the pump in the static mode with the help of an SET electrometer \cite{KellerScience99}.

In the present experiment, we address the insulation properties of
a novel single-gate electron turnstile, recently realized as a hybrid SET
transistor, built on two superconductor--insulator--normal metal (SIN) tunnel
junctions \cite{Pekola-NaturePhysics08}. In this turnstile, the interplay of
the Coulomb blockade and the sub-gap current suppression enables the accurate
transfer of single electrons, using only one gate signal. A promising estimate
was obtained theoretically \cite{AverinPekola08, Kemppinen09} for the
fundamental accuracy of such a turnstile to be at the relative level of
$10^{-8}$ for currents up to $\unit[10]{pA}$. Parallel operation of up to
10 devices driven by a common rf-gate, producing a total current above
$\unit[100]{pA}$, has been demonstrated \cite{10turnstiles}. On the other
hand, having only two junctions in-series, $N=2$, instead of, for instance \cite{KellerScience99},
$N=7$,  was found to be a critical issue for the
metrological applications due to non-vanishing sub-gap leakage currents,
deteriorating the pumping performance.

It has been predicted theoretically \cite{OdinBubaScho,GolZaik}
and proven experimentally \cite{ZorinJAP00,LotkTrap99,Lotk3jpump01}
that for few-junction normal-state SET circuits, there is an
efficient way to suppress the cotunneling leakage by engineering the
electromagnetic environment of the junctions.
Quantitatively, the implementation of an in-series high-ohmic local resistor
$R$ provides a result similar to an extension of the series array
by  $n$ additional tunnel junctions, where  $n \sim R/R_{\rm Q}$ and $R_{\rm Q}
\equiv h/e^2 \approx \unit[25.8]{k\Omega}$ is the resistance
quantum. For example, the hold times in a 4-junction R-trap (array with the resistor
$R \approx \unit[50]{k\Omega}$) on the scale of hours \cite{LotkTrap99} were
found comparable to those for 7- to 9-junction traps without resistors
\cite{LukensTrap94,KrupeninTrap96}.

Compared to the normal-state case, there is a possibility of an additional
first-order error process in the hybrid turnstile that involves tunneling
of electrons, either photon assisted \cite{PekolaEA10} or via the Dynes
density of sub-gap states \cite{Dynes}. However, even in this case,
a high-ohmic environment can be helpful for reducing the leakage. In the
previous experiment on R-turnstiles (a hybrid turnstile connected in series
to the resistors, see Ref.~\cite{Lotkhov09}), we were able to achieve more
than an order of magnitude suppression of the sub-gap current.

Rapid progress in understanding the sub-gap tunneling phenomena as
well as the efficient methods \cite{PekolaEA10,Lotkhov09},
developed for their suppression, brought the frontier of experiments
down to a single-electron level \cite{SairaBox10} and the current
uncertainty: much closer to the level demanded by metrology. In this
work, we report on the behavior of the R-turnstile in the static hold mode,
\ie~in the configuration, where it is connected to a small "trapping"
island monitored by an SET electrometer. The hold times, measured in
our experiment on the scale up to $\tau \sim \unit[20]{s}$, are
promising for metrological applications, due to exceptionally low
sub-gap current uncertainty $\Delta I \sim e/\tau \sim \unit[10^{-20}]{A}$.
The electron switchings are analyzed along the lines of
the model of environmental activation of tunneling, proposed in
previous works \cite{PekolaEA10,SairaBox10}.

\section{Fabrication and Experiment}

\begin{figure}[]
\centering%
\includegraphics[width=0.7\columnwidth]{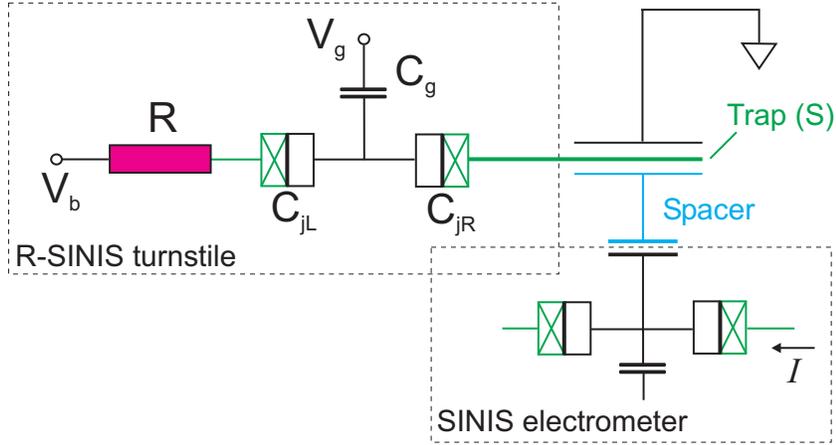}
\caption{Equivalent circuit of the sample, including an R-SINIS
turnstile and trapping island coupled capacitively via a spacer
electrode to a SINIS electrometer. The open/crossed boxes indicate
the normal/superconducting electrodes of the tunnel junction.}
\label{Circuit}
\end{figure}

\begin{figure}[]
\centering%
\includegraphics[width=0.7\columnwidth]{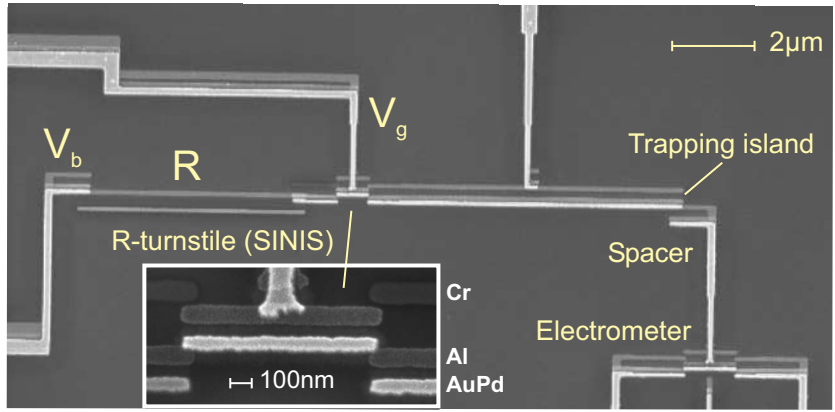}
\caption{SEM picture of the experimental structure. The inset shows
a close-up of the double-junction structure of the turnstile.}
\label{SEM}
\end{figure}

The experimental circuit is shown in Fig.~\ref{Circuit}. It includes
a hybrid SINIS-type R-turnstile, terminated by the trapping island
capacitively coupled to a SINIS SET electrometer. An SEM image of the structure
is shown in Fig.~\ref{SEM}. It was fabricated on the thermally oxidized
Si substrate by means of the standard shadow evaporation technique
\cite{Shadow}. The layout consists of three replicas of different
materials: Cr for the resistor, $\unit[12]{nm}$-thick, evaporated in
oxygen at a pressure of about $\unit[1.5 \times 10^{\rm
{-4}}]{Pa}$; Al ($\unit[18]{nm}$) oxidized just after the
evaporation at $\unit[1]{Pa}$ for $\unit[5]{min}$ to form the tunnel
barrier; and a proximity bilayer of Al ($\unit[5]{nm}$) and AuPd
($\unit[20]{nm}$), building together the normal-state island of the
turnstile. Here, the thin Al film was used in order to improve the quality
of  the tunnel barrier. For the purpose of increasing the yield of the
intact junctions, the layout included a semiconducting ring of evaporated
Si (not shown), cross-connecting all on-chip lines during sample manipulations
at room temperature \cite{NMDC2009}.

The trapping island was made of solely the superconducting
Al, to avoid unpaired electron states within the energy gap
$\Delta$. The island was coupled to the electrometer
indirectly through its capacitive link to the neighboring
normal-conducting replica, connected galvanically to a spacer
electrode, as shown in Fig.~\ref{SEM}. Due to the indirect way of
coupling, the electrometer response function was sufficiently weak
to avoid the back influence of this charge detector onto the trap (see below).

\begin{table}
\caption{\label{Sampa} Estimates of the sample parameters. Here, $R_{\rm T}=R_{\rm T}^{\rm {left}}=R_{\rm T}^{\rm {right}}$
is a junction resistance in the turnstile, $E_{\rm C} \equiv e^2/\left [2 \times \left (C_{\rm {jL}}+C_{\rm {jR}}+C_{\rm g}\right )\right ]$
the turnstile charging energy, and $\tau_{\rm {max}}$ the maximum expectation value
of the hold time.}
\begin{indented}
\lineup
\item[]\begin{tabular}{@{}lllllll}
\br
Sample No. & $R_{\rm T},~k\Omega$ & $E_{\rm C},~\mu eV$ & $\Delta_{\rm {Al}},~\mu eV$ &
$R(0),~k\Omega$ & $R(\infty),~k\Omega$ & $\tau_{\rm {max}},~s$\\
\mr

  1  & 600 & 750 & 260 & 85  & 44 & 25.0 \\
  2  & 120 & 360 & 230 & 38  & 36 & \02.6 \\
\br
\end{tabular}
\end{indented}
\end{table}

The measurements were carried out in a dilution fridge with a base
temperature of $\unit[15]{mK}$. The sample was encapsulated into a
microwave-tight, but not a vacuum-sealed sample holder.
Thermocoax$\texttrademark$ coaxial filters, approximately
$\unit[1-1.5]{m}$-long, were installed into the sample-holder for
every signal line. One exception was, however, made for the
high-frequency gate line (used in the present experiment to apply a
dc-voltage $V_{\rm g}$), equipped with a much shorter coaxial
filter, about $\unit[5]{cm}$.

The results of this paper are demonstrated on two samples
with substantially different parameters, determined with the help of
relevant test devices and summarized in Table~\ref{Sampa}. We note
that, due to high oxygen content, our resistive Cr lines exhibited
slightly non-linear $IV$-curves, characterized by the different
zero-bias and asymptotic resistances $R(0) \gtrsim R(\infty)$.

\section{Results and discussion}

\begin{figure}[]
\centering%
\includegraphics[width=0.7\columnwidth]{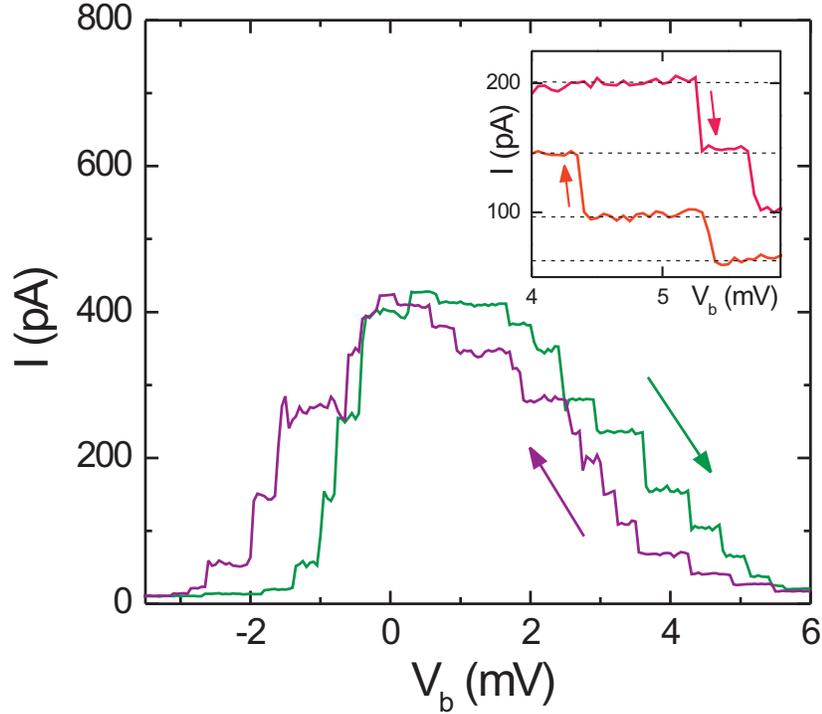}
\caption{A step-like response of the electrometer signal to the
single-quasiparticle charging in Sample~1. For this sweep, the
continuous polarization of the trapping island is permanently
canceled by a gate signal $\delta V_{\rm g} = -0.8 \times \delta
V_{\rm B}$. The trapping effect gives rise to a hysteretic
displacement of the curve, should the sweep direction be reversed.
The notable irregularity of the  state transitions appears due to the
stochastic nature of the switching processes studied in this work.
Inset: a hysteresis loop indicating quasi-stability of up to three
charge states.} \label{ladder}
\end{figure}

An important manifestation of the trapping effect is a
hysteresis, appearing in the charge detector response, as we reverse the
direction of charging the trap, see Fig.~\ref{ladder}. A
single-electron trapping induced a small polarization charge on the
electrometer island, corresponding to a $1/24$th part of the modulation
period, as shown in part by a step-like modulation function in
Fig.~\ref{ladder}. During the voltage sweep $V_{\rm b}$ in Fig.~\ref{ladder},
we applied a cross-talk cancellation signal to the gate $V_{\rm g}$, which was
used to counterbalance the continuous voltage modulation of the trapping island and
the electrometer island. As a result, the discrete detector signal was observed,
due to electron tunneling events only, as manifested in the inset to Fig.~\ref{ladder}
by a stepped shape of the hysteresis loops.

\begin{figure}[]
\centering%
\includegraphics[width=0.7\columnwidth]{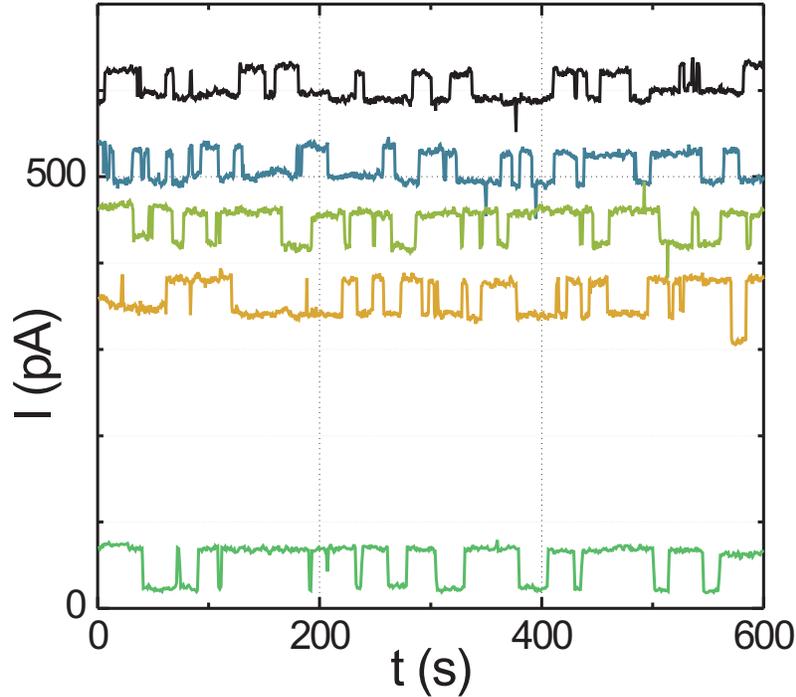}
\caption{Random state switchings in Sample~1, detected by the
electrometer, biased to five different operation points and thus
related to different levels of the dissipated power $P \equiv I
\times V_{\rm {EM}}$, $V_{\rm {EM}}$ being the bias voltage of the
detector. From bottom to top: $P \approx \unit[0.03, 0.32, 0.44,
0.53, 0.66]{pW}$. The corresponding life time expectations: $\tau =
\unit[25, 17.1, 17.6, 15.4, 19.3]{s}$, indicate only a weak
influence of the electrometer onto the trapping circuit.}
\label{backinfl}
\end{figure}

As previously mentioned, the detection of incremental charging of
the trapping island was done by means of a voltage-biased SINIS transistor.
In order to obtain a measurable output current, the bias voltage was increased
above the gap value, $V_{\rm {EM}} > 2\Delta /e$, which is sufficiently
high to enable appreciable quasiparticle current. Accordingly,
one could expect significant disturbances, imposed on the trap by the
quasiparticle recombination processes in such a detector. However, as
shown in Fig.~\ref{backinfl}, the frequency of random switchings in the
trap was not noticeably affected within the practical range of the electrometer
currents, $I \sim \unit[500]{pA}$.

\begin{figure}[]
\centering%
\includegraphics[width=0.7\columnwidth]{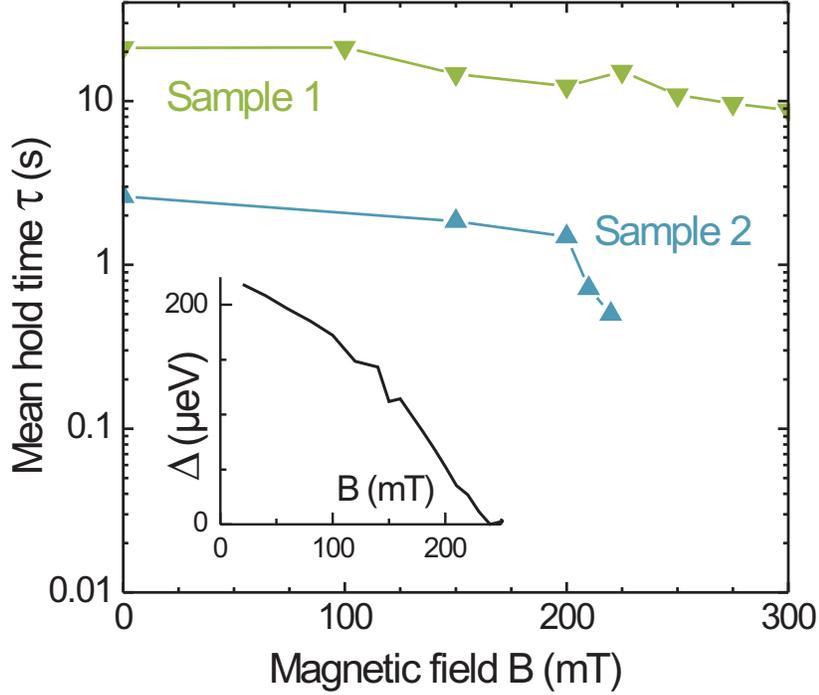}
\caption{Average hold time as a function of the magnetic
field for Samples~1 and 2. Inset: a typical dependence of the
superconducting energy gap $\Delta$ of our Al-film on the
perpendicular magnetic field $B$, measured on another sample.}
\label{taub}
\end{figure}

Since, depending on the voltage settings, both one-electron and
two-electron (Andreev reflections) tunneling events can, in principle,
contribute to the detector signal, we first calibrated the
experimental circuit, by driving the sample into the normal state
and monitoring the single-electron switchings. In Fig.~\ref{taub}
the dependence is shown of the average hold time on the magnetic
field $B$, arranged perpendicular to the substrate. The inset
illustrates a suppression of the energy gap $\Delta$ with increasing
values of $B$. The comparison with the normal-state behavior brought
us to the conclusion that the majority of events were related to
1e-tunneling even in the superconducting state. Rare 2e-events were
also observed, but they were statistically indistinguishable from a
possible small amount of time-unresolved two-electron tunneling
sequences. On the other hand, theoretically, a single-electron
escape through the SINIS turnstile must leave behind a quasiparticle
excitation in the trapping island. This process is, however,
prohibited energetically at low bias voltage and low temperatures.
Therefore, observation of predominantly single-electron events in
our experiment must be attributed to the dominance of
non-equilibrium carriers in the escape processes.

The average hold times of two equally populated neighboring charge
states were registered to be at the level of $\tau \approx \unit[20]{s}$,
in the high-$E_{\rm C}$ energy Sample~1, and $\tau \approx \unit[2.6]{s}$,
in the moderate-$E_{\rm C}$ Sample~2. These values are remarkable, because
such long hold times have been previously observed only for
single-electron arrays without resistors with the number of junctions
$N \ge 4$ (see Refs.~\cite{FultonTrap91,LafargeTrap91,Martinis5jpump94}).

\begin{figure}[]
\centering%
\includegraphics[width=0.7\columnwidth]{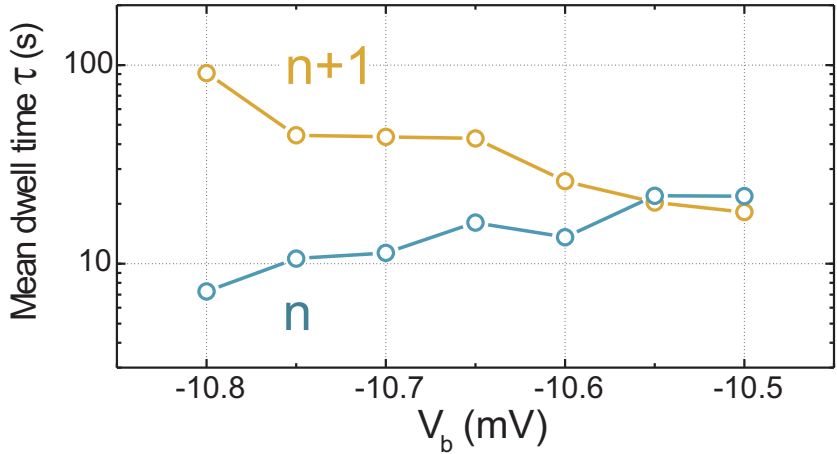}
\caption{Average hold time as function of voltage $V_{\rm b}$,
measured for two neighboring states with $n$ and
$n+1$ trapped electrons. The point $V_{\rm {b0}} \approx \unit[-10.55]{mV}$
corresponds to a center of the hysteresis loop (a zero-bias point).
The points to the left of this symmetry point depict the hold times in the finite-bias states
of the turnstile.} \label{taubias}
\end{figure}

Due to the gate modulation by the cancellation signal $\delta V_{\rm g}$ (see above),
a strong systematic variation of the loop width was observed, as we
swept over more than $\sim 10$ loops. However, for the hold times
measurements, we focused on the loop of the maximum width,
where only a little change to the neighboring loops was observed.
On the other hand, due to the same cancellation signal, an effective bias
voltage across the R-SINIS-turnstile-part of the circuit, which was zero
in the center of the loop $V_{\rm {b0}}$, could be directly obtained
in experiment as $\left(V_{\rm b}-V_{\rm {b0}}\right)$  within
the same loop. The latter consideration enables a direct estimate of
the practical accuracy of the turnstile at finite bias voltage with
the help of the data shown in Fig.~\ref{taubias}. One can see that for the
half of the gap voltage, which typically corresponds to an optimal bias in
the center of the pumping plateau \cite{Pekola-NaturePhysics08,Lotkhov09},
the electron leak is still on the scale of one electron per $\unit[10]{s}$.
Accordingly, for the R-turnstile driven by a rectangular signal \cite{Kemppinen09},
\ie~ flipped quickly between two optimized gate settings, one could expect a
leakage-limited accuracy, say, at the pumping frequency $f = \unit[10]{MHz}$,
on the metrological level $10^{-8}$.

\begin{figure}[]
\centering%
\includegraphics[width=0.7\columnwidth]{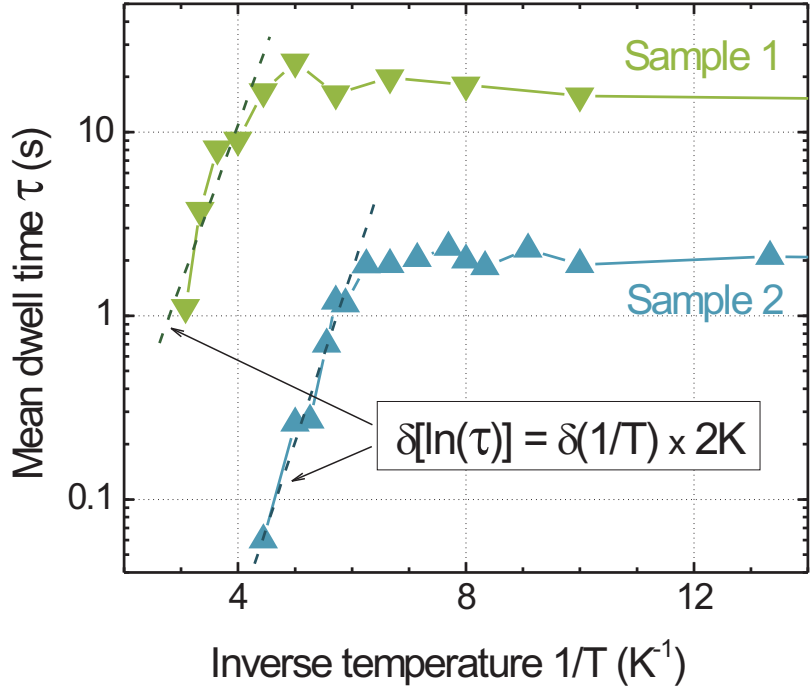}
\caption{ Average hold time as a function of the inverse temperature
for Samples~1 and 2. A rough comparison with the Ahrrenius plot
reveals qualitatively no correlation of the high-temperature slopes
with $E_{\rm C}$ (differing for these samples by as much as a  factor
of two). See text for further details.} \label{taut}
\end{figure}

Furthermore, our samples demonstrated no deterioration of the hold time $\tau$
within wide temperature ranges up to $\unit[150...200]{mK}$ (see Fig.~\ref{taut}).
This predicts excellent stability against the electron overheating effects
for pumping currents up to $\sim \unit[100]{pA}$ \cite{Maibaum}. At higher
temperatures, a fast exponential reduction of the hold times was observed.
A slope of the decay, estimated in Fig.~\ref{taut} to be similar for both
samples, reveals a phenomenological activation energy of about
$\left( k_{\rm_B} \times \unit[2]{K} \right)$, which is considerably lower
than $\Ec+\Delta$ in either sample. Our numerical analysis, which we address below,
indicates that these deviations from the simple Ahrrenius model of an
autonomous SET trap, which basically assumes a strong dependence of $\tau$ on
$\Ec$ (cf., e.g., Refs.~\cite{LukensTrap94,KrupeninTrap96}), appears due to the
influence of the on-chip resistor and the shape of the BCS density of states
in the superconductor.

\section{Model of environmental activation of tunneling for the R-SINIS trap}

In order to facilitate the understanding of the observed escape processes,
we have applied a semi-phenomenological model of environmentally enhanced
tunneling in the R-SINIS trap, which assumes that a considerable
noise contribution arrives from the warmer parts of the
electromagnetic circuitry. Further details related to this Environmental Activation
of Tunneling (EAT) approach can be found elsewhere \cite{PekolaEA10,SairaBox10}.

\subsection*{EAT formalism}

Our modeling of electron escape from the trap is based on the environmental
activation of single-particle tunneling events. The theoretical framework
is that of standard $P(E)$ theory \cite{IngoldNazarov}, which accounts
for the photon exchange with the tunnel junction, due to the excitation of
the environmental modes. Our model of the electromagnetic environment
of the trap junctions includes two independent fluctuators, as shown
symbolically in Fig.~\ref{fig:circuit_onchipr}:

\begin{flushright}
\begin{description}
\item [Fluctuator~1] represents our well-characterized on-chip chromium resistor,
which we model as an $RC$ transmission line at the sample stage temperature.
A similar case: a first-order tunneling in the presence of on-chip resistors,
has been modeled successfully with $P(E)$ theory in earlier work
\cite{Farhangfar_ExprOnTunneling,Farhangfar_EffectOfEMEnv}.

\item [Fluctuator~2] accounts for the high-frequency noise originating
possibly from a source external to the sample stage and having higher
temperature. Activation of first-order tunneling by high-temperature
environment, $\kB \Tenv \gtrsim \Delta+\Ec$, has been previously shown
to reproduce the experimentally observed finite sub-gap conductance of
NIS junctions \cite{PekolaEA10}, and low-temperature saturation of tunneling
rates in an NIS single-electron box\cite{SairaBox10}.
\end{description}
\end{flushright}

In the framework of $P(E)$ theory\cite{IngoldNazarov}, a single-electron
tunneling rate is derived on the basis of the absorption/emission spectrum
of the environment, the function $P(E)$. For Gaussian fluctuators, $P(E)$
can be evaluated as
\begin{equation}
P(E) = \frac{1}{2\pi\hbar} \int_{-\infty}^\infty \D t\,\exp \left(J(t)
+ \frac{i}{\hbar} E t\right) \label{eq:pe},
\end{equation}
where $J(t)$ is a linear phase-phase correlation function, introduced in
Ref.~\cite{IngoldNazarov} to describe fluctuations of voltage across
the tunnel junction. In the case of a linear environment in
thermal equilibrium, which can be characterized by an impedance $Z(\omega)$,
connected in parallel to the junction of capacitance $\Cj$, $J(t)$ equals
\begin{equation}
\fl \qquad J(t) = 2 \int_0^\infty \frac{\D \omega}{\omega} \frac{\Re
Z_t(\omega)}{R_K} \times \left\{\coth\left(\frac{1}{2}\beta\hbar\omega\right)\left[\cos(\omega
t)-1\right]-i\sin(\omega t)\right\},\label{eq:Jt_Zt}
\end{equation}
where $Z_t(\omega) = \left[Z^{-1}(\omega) + i\omega \Cj\right]^{-1}$
is the total im\-ped\-ance seen over the junction and
$\beta \equiv \left(\kB T_\txt{env} \right)^{-1}$ is the inverse temperature of
the environment. In order to describe the experimental setup, we will generalize
the above result to the two-fluctuator model, where the voltage fluctuations
caused by the independent fluctuators are additive. Consequently, the total $P(E)$
function can be found as a convolution
\begin{equation}
P(E) = \int\limits_{ - \infty }^{ + \infty } {dE'P_1 (E - E')P_2 (E')}
\label{eq:conv}
\end{equation}
of the partial $P_{\rm {1,2}}(E)$ functions, evaluated independently for
the two components (cf., \eg, Ref.~\cite{IngGrabEberh94}).

For a description of filtering and voltage division effects, we allow for an
arbitrary transfer function $A(\omega)$, which relates the fluctuating current
$\IN(\omega)$ through the impedance $Z(\omega)$ to the voltage $V(\omega)$ across
the junction, so that $\IN(\omega) = A(\omega) V(\omega)$. The actual phase
fluctuations across the junction will then be equivalent to those across an
effective total impedance
\begin{equation}
Z_{t}^{-1}(\omega) = A(\omega) \frac{\Re \left[ A(\omega) \right]}{\Re
\left[ Z^{-1}(\omega) \right]},\label{eq:Zt_eff}
\end{equation}
which can be used for calculations of the $P(E)$ function with the help of equations \ref{eq:pe},\ref{eq:Jt_Zt}.

\subsection*{Fluctuator 1: On-chip resistor}

\begin{table}
\caption{\label{tbl:sample_par}Sample parameters related to Fluctuator~1 and the inferred capacitive voltage division factors $\xi_\txt{L(R)}$.}
\begin{indented}
\lineup
\item[]\begin{tabular}{@{}llllllllllll}
\br
Sample & $l, \mu m$ & $R,~k\Omega$ & $C, aF$ & $C_{\rm {jR,L}}, aF$ & $C_{\rm g}, aF$ & $C_{\rm x}, aF$ & $C_{\rm t}, aF$ & $\xi_{\rm L}$& $\xi_{\rm R}$\\
\mr
1 &5&85$^{\rm a}$&320&40&30&100&24&0.58&0.29\\
2 &5&38&320&100&30&100&44&0.44&0.28\\
\br
\end{tabular}
\item[] $^{\rm a}$Zero-bias resistance $R(0)$ was used instead of the asymptotic value $R(\infty)=\unit[44]{k\Omega}$.
\end{indented}
\end{table}

\begin{figure}
\centering%
\includegraphics[width=0.7\columnwidth]{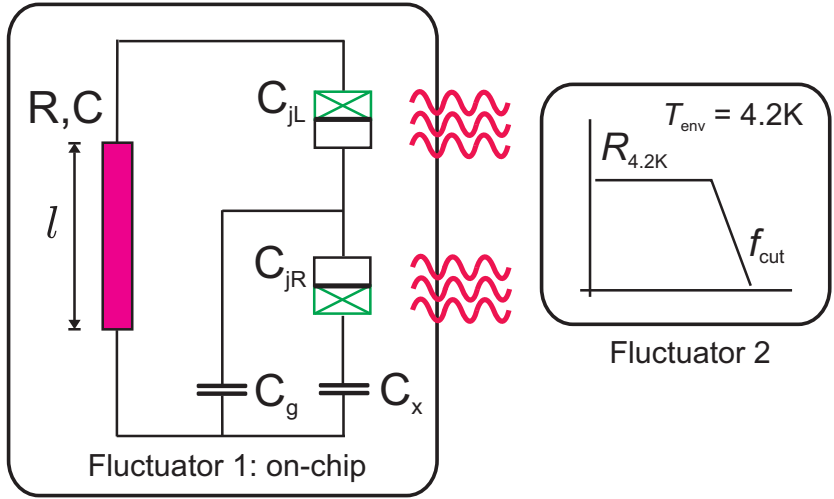}
\caption{Schematic representation of the two-fluctuator model used in simulations.
The circuit diagram to the left was used for evaluating the $P(E)$-function related
to the on-chip resistor (Fluctuator~1). The right-hand part of the plot shows
symbolically a "warmer" Fluctuator 2.}
\label{fig:circuit_onchipr}
\end{figure}

For evaluating $P_1(E)$ corresponding to the on-chip resistor, we
will use the reduced circuit diagram, shown in detail in the
left-hand part of Fig.~\ref{fig:circuit_onchipr}. $R$ and $C$
denote the total capacitance of the Cr wire, and $r = R/l$ and
$c = C/l$ where $l$ is the wire length. For $c$, we take a
rule-of-thumb value $c = 6\epsilon_0 = 5.3 \times 10^{-11}\unit{F/m}$
accounting for geometric stray capacitance. We have introduced a shorthand
notation $A || B := \left( A^{-1} + B^{-1} \right)^{-1}$ for impedance
calculations.
The essential $RC$ transmission line parameters are the frequency dependent
wave number, $\alpha(\omega) = (1+i)\sqrt{\frac{\omega r c}{2}}$, and
the characteristic line impedance, $Z_0(\omega) = (1-i)\sqrt{\frac{r}{2 \omega c}}$.
For a semi-infinite line, the impedance seen from the open end equals the
characteristic impedance, and for a finite wire terminated with a short circuit,
one finds $Z(\omega) = Z_0(\omega) \tanh\left[\alpha(\omega) l\right]$.

Next, we note that the electrical circuit in parallel to the
resistor consists of purely capacitive elements only. Hence, it can be
modeled by a single lumped capacitor of value $C_t = C_\txt{jL} ||
(C_\txt{g} + C_\txt{jR} || C_\txt{x})$. The voltage fluctuations over
the junctions $L$ and $R$ are equivalent to the fluctuations over $C_t$,
scaled by a factor $\xi$ accounting for the capacitive voltage division.
For the left (upper) junction, we have $\xi_\txt{L} = \left[1 + C_\txt{jL}/(C_\txt{g} +
C_\txt{jR}||C_\txt{x})\right]^{-1}$, and for the right,
$\xi_\txt{R} = \left(1 + C_\txt{jR}/C_\txt{x}\right)^{-1} \left[1 + (C_\txt{g} +
C_\txt{jR}||C_\txt{x})/C_\txt{jL}\right]^{-1}$. According to
\Eref{eq:Zt_eff}, voltage division by factor $\xi$ is
equivalent to scaling $Z_t(\omega)$ by a factor of $\xi^2$. Based on
the preceding argument, we can now write the effective impedance,
used for evaluating the $P_1(E)$ function related to Fluctuator 1, as
\begin{equation}
Z_t(\omega) = \frac{\xi^2}{i \omega C_t +
Z_0^{-1}(\omega) \coth\left[\alpha(\omega) l\right]}.\label{Zt_onchip}
\end{equation}

The numerical values of the model parameters are presented
in Table~\ref{tbl:sample_par}.

\subsection*{Fluctuator 2: A source of high-frequency noise}

The principal escape mechanism is assumed to be related to the
excitation of quasiparticles by the photons with energy
$E_{\rm {ph}} \sim \Delta + \Ec$, which is high enough to overcome the
Coulomb blockade in the turnstile. For both studied samples
$\Ec > \Delta$, and hence the relevant noise frequencies are
above $f_{\rm {ph}} \equiv E_{\rm {ph}}/h  \sim \unit[100]{GHz}$.
The photon excitations are expected to arrive from the warmer (and more
distant) parts of the environment, to be modeled as Fluctuator 2, shown
symbolically in the right-hand part of Fig.~\ref{fig:circuit_onchipr}.

Due to the lack of microscopic understanding, we will model Fluctuator~2
just as a parallel $RC$ circuit at $T = \unit[4.2]{K}$
(see previous work in Refs.~\cite{PekolaEA10,SairaBox10}),
coupled directly to either junction. Used as a model parameter, the
cut-off frequency $f_{\rm {cut}}$ of this effective circuit can be reasonably
interpreted as the cut-off of the coupling between the source and the junction.
Taking into account that the gate capacitances in the reported turnstiles are
comparable to those of the (very small) tunnel junctions, we assume the gate
lines are the dominant on-chip propagation ways for the external noise.
(For simplicity, this concern was disregarded in Fig.~\ref{fig:circuit_onchipr}.)
On the contrary, the high-frequency power attenuation for the on-chip Cr resistors
is very strong: if modeled as $RC$ transmission lines, it takes the exponential
form $\exp(-\sqrt{2 \omega R C})$ with the cut-off frequency $1/(2\pi RC) \sim \unit[10]{GHz} << f_{\rm {ph}}$.

As long as the hold time at low temperatures depends mainly on the
spectral density of environmental emission at high frequencies $\sim~f_{\rm {ph}} \gg f_{\rm {cut}}$,
the source resistance and cutoff frequency of the high-temperature environment
cannot be independently extracted from the measurements. In the calculations,
a constant cut-off frequency $f_{\rm {cut}} = \unit[0.5]{GHz}$
was used, corresponding to the filtering properties of the Thermocoax$\texttrademark$
lines. We stress that the effective source resistance $R_\txt{4.2K}$,
obtained as a model fit parameter, formally includes the details of a phenomenological
transfer function $A(\omega)$ and is not expected to be equal to the actual source
resistance.

\subsection*{Results}

\begin{figure}
\centering%
\includegraphics[width=0.7\columnwidth]{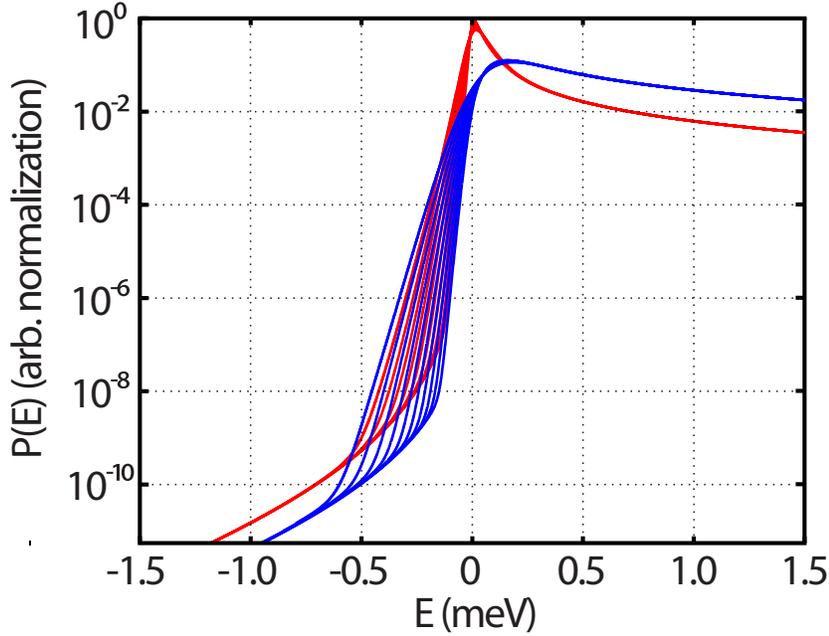}
\caption{$P(E)$ curves calculated with respect to the junctions L
(blue curves) and R (red curves), for the parameters of Sample~1
(see Table~\ref{tbl:sample_par}) and various temperatures in the
range 50--300~mK (in the order of broadening of the peak towards the
higher temperatures).}
\label{fig:PE_example}
\end{figure}

The typical composite $P(E)$ functions, resulting from the calculations
described above, are shown in Fig.~\ref{fig:PE_example}. The slowly
decaying environmental absorption spectrum ($E > 0$) is due to the on-chip
resistor, whereas the finite emission probabilities on the millivolt scale at
low temperatures stem from the high-frequency noise. A maximum of
the $P(E)$ functions occurs at a positive voltage due to the dynamic
Coulomb blockade arising owing to the on-chip resistor.

The single-electron tunneling rates were computed using the golden
rule formula
\begin{equation}
\fl \qquad \Gamma(U) = \frac{1}{e^2\RT}\int_{-\infty}^\infty dE_1\int_{-\infty}^\infty dE_2
\nS(E_1) \fS(E_1) [1-\fN(E_2)] P(E_1-E_2+U),\label{eq:goldenrule}
\end{equation}
where $\RT$ is the junction resistance, $\nS(E)$ the density of
states in the S electrode, and $f_\txt{N(S)}$ the occupation factor for
quasiparticle states in the N(S) electrode. In our model, we use the
pure BCS density of states $\nS(E) = |\Re \frac{E}{\sqrt{E^2 -
\Delta^2}}|$ and the occupation factors are taken to obey equilibrium
Fermi distribution, \ie, $f_{N,S}(E) = \left[1 + \exp(E/\kB T)\right]^{-1}$,
where $T$ is the sample stage temperature. The assumption of the equilibrium
quasiparticle distribution in the S electrodes is based on the ultra
low injection rates (of the order of $\unit[1]{Hz}$), allowing for
sufficiency of the quasiparticle relaxation/recombination rates
on both sides of the turnstile. The tunneling rates $\Gamma_\txt{L(R)}$
were evaluated independently for the left and right tunnel junctions,
respectively.

Finally, the hold times $\tau$ were calculated based on the master equation
\cite{AverinLikharev91} for single-electron tunneling
\begin{equation}
\tau^{-1} = \GR(-\Ec) \frac{\GL(\Ec)}{\GL(\Ec) + \GR(\Ec)}
 + \GL(-\Ec) \frac{\GR(\Ec)}{\GL(\Ec) + \GR(\Ec)}.\label{eq:lifetime}
\end{equation}
We note that the sequences, where an electron tunnels to the turnstile
island and then back, are too fast to be detected, and hence they do
not affect the observed hold times. Furthermore, the charging energy of
the trapping node is considered to be compensated in the experiment by an
appropriate bias voltage $V_{\rm b}$, and it was not included in the
simulations.

\begin{figure}[tbh]
\centering%
\includegraphics[width=0.7\columnwidth]{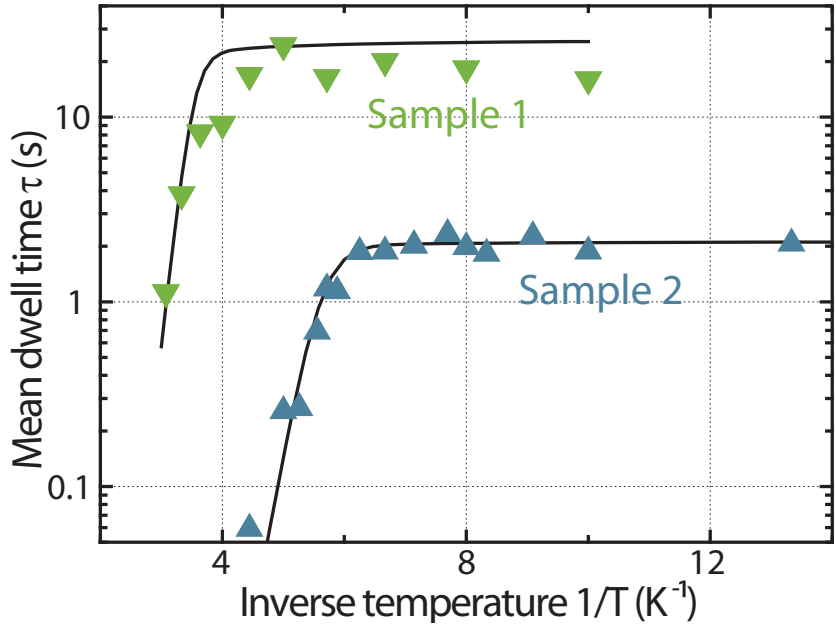}
\caption{The experimental hold times (filled symbols, shown previously
in Fig.~\ref{taut}) versus the results of the two-fluctuator model
(black curves).}
\label{fig:ahrrenius}
\end{figure}

\begin{table}
\caption{\label{tbl:para}Sample parameters accepted for the final evaluation of
the hold times in Fig.~\ref{fig:ahrrenius}.}
\begin{indented}
\lineup
\item[]\begin{tabular}{@{}ccccc}
\br
Sample & $\RT, k\Omega$ & $\Delta, \mu eV$ & $\EC, \mu eV$ & $R_\txt{4.2K}, \Omega$\\
\mr
1 &600&260&600&11.3\\
2 &120&230&360&\01.8\\
\br
\end{tabular}
\end{indented}
\end{table}

Figure~\ref{fig:ahrrenius} demonstrates a reasonable agreement of the
measured and calculated hold times in the trap. The parameters used in
numerical calculations are summarized in Table~\ref{tbl:para}. The fit value of
$\Ec = 600\unit{\mu eV}$ in Sample~1 is lower than its experimental estimate
$\Ec \approx 750\unit{\mu eV}$, which we attribute to an effect of a not quite optimum
gate voltage setting in measurements. The fit value of $R_\txt{4.2K}$ for Sample~1
was found larger than that of Sample~2 by a factor of 6.3, corresponding to a factor of
$\sqrt{6.3} \approx 2.5$ in the voltage fluctuation amplitude. Larger voltage fluctuations
can be, at least partially, attributed to the lower junction capacitance of Sample~1
compared to Sample~2 (see Table~\ref{tbl:sample_par}). Furthermore, since the photon
energy $E_{\rm {ph}}$ required to overcome the Coulomb blockade barrier is different
for the two samples, any deviation of the true noise spectrum from the assumed
first-order decay could also result in different fit values for $R_\txt{4.2K}$.

At higher temperatures, the hold times are influenced by the thermal broadening of
the Fermi-factors, but also by an increased probability of photon emission by the
on-chip resistor (see the plot $P$ vs. $E$ for $E < 0$ in Fig.~\ref{fig:PE_example}).
For Sample~1, we evaluated the hold time also in the case of a magnetically suppressed
energy gap $\Delta = 0$. The obtained value of $\unit[4.5]{s}$ is to the order of
magnitude consistent with the experimental hold time $\tau \sim \unit[10]{s}$
(see Fig.~\ref{taub}), which provides one more proof of validity for our model.

\section{Conclusions and outlook}

Our study of the hold times in the two-junction hybrid R-traps
gives rise to a promising prediction of the leakage-limited pumping
accuracy of hybrid R-turnstiles. The hold times on the scale of up
to tens of seconds imply that the metrological accuracy is feasible
and could, in principle, be realized with the help of a special
rectangular drive \cite{Kemppinen09}. A more detailed study is under
way to characterize the leak in the gate-open state of the turnstile,
which is relevant to a more common sinusoidal driving signal.
The experimental switchings have been successfully interpreted in terms
of the environmentally-induced escape processes. In particular, a good
agreement has been achieved in the frame of a two-component model of the
environment. Significant disturbances have been shown to arrive from the
warmer parts of the set-up, supposedly from outside the sample stage.
In this respect, an important further insight can be provided
by comparing the hold times of the same trapping device, measured in
different measurement setups or by varying the quality of noise filtering.

\section*{Acknowledgements}

Fruitful discussions with S~Kafanov and V~Bubanja are gratefully
acknowledged. Technological support from T~Weimann and V~Rogalya
is appreciated. The research conducted within the EU project SCOPE
received funding from the European Community's Seventh Framework
Programme under Grant Agreement No. 218783.

\end{document}